\documentclass[%
 reprint,
 amsmath,amssymb,
 aps,
]{revtex4-1}

\usepackage{graphicx}
\usepackage{dcolumn}
\usepackage{bm}

\begin{document}

\preprint{APS/123-QED}

\title{Solitons in Weakly Non-linear Topological Systems: Linearization, Equivariant Cohomology and K-theory}

\author{Daniel Sheinbaum}
 \affiliation{Facultad de Ciencias, Universidad Nacional Aut\'onoma de M\'exico. Circuito Exterior S/N, Cd. Universitaria, Colonia Copilco el Bajo, Delegaci\'on Coyoac\'an, 04510, Ciudad de M\'exico, M\'exico}
 \email{dshein@ciencias.unam.mx}

\date{\today}

\begin{abstract}
 There is a lack of knowledge about the topological invariants of non-linear $d$-dimensional systems with a periodic potential. We study these systems through a classification of the linearized NLS/GP equation around their soliton solutions. Stability conditions under linearized (mode) adiabatic evolution can be interpreted topologically and we can use equivariant $\mathit{K}$-theory and cohomology for their classification. On a lattice with crystallographic point group $P$, modes around stable, $P$-symmetric solitons are coarsely classified by the groups $\bar{\mathit{K}}_{P}^{0,\tau}(\mathbb{T}^d)\oplus H^2(BP;\mathbb{Z})$. Similarly, for $P$-symmetric gap solitons that are oscillatory stable, we have $\bar{\mathit{K}}_{P}^{0,\tau}(\mathbb{T}^d)\oplus \tilde{R}(P)$ instead. If we include a boundary, we can replace $\bar{\mathit{K}}_{P}^{0,\tau}(\mathbb{T}^d)$ with $\mathit{K}^{-1}_{P}(\mathbb{T}^{d-1})$. Finally, we mention how to use these, and the spaces of soliton solutions $M_{D}(E_{gap})$ and $M_{O}(E_{gap})$ to provide global invariants for the system.
\end{abstract}

\maketitle

\section{Introduction}\label{sec:Introduction}

The field of topological systems in photonics \cite{Top-photonics}, exciton-polaritons \cite{Exciton-Polariton} and Bose-Einstein Condensates (BEC) \cite{BEC} has brought to the fore the interplay between non-linear effects such as solitons \cite{Nonlinear-Waves}, \cite{Nonlinear-optics} and linear topological phases of matter \cite{TKNN}, \cite{Kane-Mele-2}, \cite{Avron-Seiler-Simon}. An interesting question arises when we couple a non-linear medium to a periodic potential \cite{Weinstein-Physics}, particularly when the linear system is topologically non-trivial, e.g. quantum Hall effect (QHE), SQHE and topological insulators (TI) \cite{Top-photonics}. Are systems which combine both still topological? And if so, which topological invariants classify these new non-linear systems? Though there are already many theoretical and experimental results on the stability of solitons, non-linear Bloch waves and their interactions as well as resulting macroscopic properties of materials \cite{Nonlinear-top-photonics}, \cite{Top-BandGap-Solitons}, \cite{Exciton-Polariton}, there are not many proposals for topological invariants classifying these systems (see \cite{Nonlinear-Berry}, \cite{Nonlinear-Dirac-Cones} for some results in this direction). For weakly non-linear systems, the Non-linear Schrödinger/Gross-Pitaevskii equation (NLS/GP) \cite{Nonlinear-optics}, \cite{Nonlinear-top-photonics}, \cite{Weinstein-Book} and variations of these are good approximations to low-energy behaviour. Here we shall consider weakly non-linear $d$-dimensional systems with a periodic potential \cite{Weinstein-Physics} and magnetic field described by
\begin{equation}\label{eq:NLS/GP}
    i\partial_{z}\Psi(\vec{x},z) = [-(i\nabla - \vec{A}(\vec{x}))^2 + V(\vec{x}) - f(\vec{x},|\Psi|^2)]\Psi(\vec{x},z),
\end{equation}
where $z$ can be either time (BEC) or the distance along the direction of propagation (Optics, here $z\geq 0$). We also assume that $\vec{A}(\vec{x}+\vec{a}) = \vec{A}(\vec{x});\,\,V(\vec{x}+\vec{a}) = V(\vec{x})$ for all $\vec{a} \in \Lambda$, where $\Lambda$ is a $d$-dimensional lattice. Note that in general the assumption is not true for the magnetic potential $\vec{A}$ and we should include disorder \cite{Prodan-Schulz-Baldes}. We denote the linear part by $\mathcal{H}_{l} = -(i\nabla - \vec{A})^2 + V$. We can simultaneously include systems with a boundary in our discussion by splitting $\vec{x}= (x_{\perp},\vec{x}_{||})$ and making $V(\vec{x}) = A(\vec{x})= 0$ for $ \,x_{\perp} \leq 0$ and periodic only in the $\vec{x}_{||}$-direction with respect to $\Lambda_{||}$, a $d-1$-dimensional lattice, parallel to the boundary. We would like $\mathcal{H}_{l}$ to correspond to a linear topological system (phase) with a gap \cite{Avron-Seiler-Simon} or gapped bulk condition (systems with boundary) \cite{AASS} under adiabatic evolution \cite{Avron-Adiabatic} at a fixed energy scale $E_{gap}$, as there is no analogue of the Fermi energy $E_{F}$ for our non-linear systems, since they are generally bosonic \cite{Top-photonics}, \cite{DeNittis-Maxwell}.

$\mathit{K}$-theory \cite{Hatcher-K-theory} and cohomology \cite{Hatcher-Alg-Top} have by now been used in different fields of physics, for example $\mathit{K}$-theory was used in \cite{Kane-Mele-2}, \cite{Horava-Fermi} and cohomology in \cite{Wen-PRL}, and \cite{Kapustin-PRL}, among many others. For those readers who are not familiar with algebraic topology, we present a brief description of what it entails. Given a topological space $X$, cohomology and $\mathit{K}$-theory assign abelian groups $H^n(X;\mathbb{Z})$ and $\tilde{\mathit{K}}^{0}(X)$ to it, hence the name algebraic topology. A homotopy between two maps from a topological space $X$ to $Y$ is simply a continuous family of maps parametrized by $s$ in $[0,1]$ such that at 0 and 1 it coincides with the original ones. Homotopies are the backbone of algebraic topology. Said differently, everything in algebraic topology (and topological phases) has an expression in terms of homotopy theory, e.g. cohomology, $\mathit{K}$-theory and their equivariant versions \cite{Freed-Moore} are defined using homotopy theory, the all-encompassing framework. In particular, the groups assigned to $X$, $H^{n}(X;\mathbb{Z})$ and $\tilde{\mathit{K}}^{0}(X)$ can be constructed out of maps (up to homotopy) from $X$ to another space called the classifying space, which is independent of $X$, but does depend on whether it is $H^n(X;\mathbb{Z})$ or $\tilde{\mathit{K}}^{0}(X)$.  Here, as it is done in the physics community, we interpret adiabatic evolution as a homotopy \cite{Avron-Adiabatic} and show how these classifying spaces arise from physical and stability conditions. For our purposes, equivariant topology can be thought of in the following way: If we have symmetries represented by a group $P$, we only want to consider maps that preserve them and see which groups we get with this restriction. We represent these as $H^{n}_{P}(X;\mathbb{Z})$ and $\mathit{K}_{P}^{0}(X)$ respectively. With this in mind let us proceed.

For fully periodic systems with a gap, there is a coarse classification \footnote{Coarse means we do not care about adding trivial valence bands to our systems.} where a phase $[\mathcal{H}_{l}-E_{gap} I] \in \tilde{\mathit{K}}^{0}(\mathbb{T}^{d})$, the $\mathit{K}$-theory group constructed out of vector bundles over the Brillouin torus $\mathbb{T}^{d}$ arising from the Bloch bands below the Fermi energy \cite{Kitaev}, \cite{Freed-Moore}. Meanwhile, for systems with a boundary, $\mathit{K}$-theory arises naturally and $[\mathcal{H}_{l}-E_{gap}I]\in \mathit{K}^{-1}(\mathbb{T}^{d-1})$, where $\mathbb{T}^{d-1}$ is the surface Brillouin torus \cite{AASS}. Ideally we would imitate the linear classification for systems with $f(\vec{x},|\Psi|^2)$ by defining a non-linear gap condition and a notion of non-linear adiabatic evolution \cite{Nonlinear-Adiabatic}. The problem is that so far there is no analogue of a gap condition for non-linear systems \cite{Nonlinear-top-photonics}. We will consider the simplified problem of classifying the topological behaviour of modes around soliton solutions, stationary solutions of the form $\Psi(\vec{x},z) = e^{-i\lambda z}\Phi^{\lambda}(\vec{x})$ to eq (\ref{eq:NLS/GP}), which decay exponentially as we go to spatial infinity.

But, can all solitons have topological modes? And even if some do, could these be destroyed by the non-linearity? If a soliton is unstable it will eventually disappear and its modes together with it. Positive (ground state) solitons have two types of instabilities, one is a \textit{focusing} instability \cite{Weinstein-Physics}, where without increasing, its energy  focuses towards a single point, yielding an arbitrary high density which  blows up. The other is a \textit{drift} instability, where via asymmetric distortions, infinitesimal displacements of its original position make it drift towards infinity \cite{Weinstein-Physics}. What happens for non-positive solitons, so-called gap solitons, i.e. those which live between the spectral bands of the linear problem? These suffer from a different set of instabilities \cite{Pelinovsky-Oscillatory}. One of these is an \textit{oscillatory} instability, where among other mechanisms, a vibration mode appears and resonates with radiation, causing the soliton to oscillate at higher and higher frequencies \cite{Pelinovsky-Oscillatory-Numerical}. We would like our solitons to avoid such instabilities; therefore, we shall further impose some stability conditions \cite{Weinstein-Physics}, \cite{Pelinovsky-Oscillatory}. However, how compatible are these conditions with those necessary for topological modes? Do these conditions have a topological interpretation? We shall elucidate their topological character and their relation to the other topological restrictions in what follows.

\section{Linearization and stability conditions}\label{sec:Linearization}
We linearize eq (\ref{eq:NLS/GP}) around $\Phi^{\lambda}$ \cite{Weinstein-Zhou}, \cite{Weinstein-Physics} which yields
\begin{equation}\label{eq:LNLS}
    \partial_{z}\vec{\chi} = \mathcal{L}(\lambda)\vec{\chi},
\end{equation}
where 
\begin{equation} \label{eq:bigL}
    \mathcal{L}(\lambda) =
    \begin{pmatrix}
    0 & L_{-}(\lambda)\\
    -L_{+}(\lambda) & 0
    \end{pmatrix},
\end{equation}
with self-adjoint operators
     \begin{eqnarray}\label{eq:L-L+}
    L_{-}(\lambda) &=& \mathcal{H}_{l} -\lambda I -f(x,|\Phi^{\lambda}|^2) ,\\
     L_{+}(\lambda) &=& \mathcal{H}_{l} -\lambda I -f(x,|\Phi^{\lambda}|^2) - 2df|_{\phi^{\lambda}}. \nonumber
\end{eqnarray}
Our linearized problem has a clear analogue of a gap condition for the mode operators $L_{\pm}(\lambda)$. We note that $\mathcal{H}_{l} -\lambda I$ has its spectrum shifted and hence the scale at which the soliton satisfies a \textit{gapped modes} condition is at
\begin{eqnarray}\label{eq:mode-gap}
    E_{modes}(\lambda) &=& E_{gap}-\lambda, \\
    0 &<&\lambda < E_{gap}.
\end{eqnarray}
Thus, the first constraint we put on solitons to have topological modes is $\lambda < E_{gap}$.
We also have extra potentials determined by $f(x,|\Phi^{\lambda}|^2)$ and $f(x,|\Phi^{\lambda}|^2) -2df|_{\phi^{\lambda}}$, which we name the \textit{soliton potentials}. For the soliton potentials not to destroy the topological character of $\mathcal{H}_{l}- \lambda I - E_{modes}(\lambda)$, we need them to behave as a perturbation/impurity/defect that does not break the gapped modes condition
\begin{eqnarray}\label{eq:Soliton-potential}
  |f(x,|\Phi^{\lambda}|^2)| &<<& E_{modes}(\lambda),\\
  |f(x,|\Phi^{\lambda}|^2)-2df|_{\phi^{\lambda}} &<<& E_{modes}(\lambda).\label{eq:Soliton-potential2}
\end{eqnarray}
These constrain both the non-linearities and the width of solitons with topological modes. Just as an example, for power non-linearities $f(|\Psi|) = |\Psi|^{p-1},\,\,p>1$, solitons which have a peak of the order $E_{modes}(\lambda)^{\frac{1}{p-1} }$ will destroy the gapped modes condition and this will be easier as $\lambda$ increases. For systems with a boundary the constraint is only necessary for solitons that are surface-localized. Here arises our first connection with instabilities. Positive (ground-state) solitons that are focusing unstable \cite{Weinstein-Physics}, that is, solitons whose modulus blows up will quickly break the gapped modes condition. This means that our solitons should satisfy the Vakhitov-Kolokolov stability condition
\begin{equation}\label{eq:VK}
    \frac{dP}{d\lambda}< 0,
\end{equation}
where $P = \int |\Psi|^2d\vec{x}$ is the particle number (BEC) or optical power, which is conserved. Let us now consider solitons that are drift stable, i.e. those which stay put under small displacements of their initial position. These have to satisfy the spectral condition
\begin{equation}\label{eq:SpectralCondition}
    n_{-}(L_{+}(\lambda)) = 1,
\end{equation}
where $n_{-}(L_{+}(\lambda))$ is the number of negative eigenvalues. Note that for positive solitons, these conditions are necessary and sufficient for full stability \cite{Weinstein-Physics}. The latter condition can be interpreted as topological if we further note that from the positivity of $\mathcal{H}_{l}$, there is a restriction on the continuous spectrum $\sigma_{c}(L_{+}(\lambda))$ (scattering modes) to be positive. Modes associated to $L_{+}(\lambda)$ live in a Hilbert space $\mathfrak{H}^{2}(\mathbb{R}^{d},\mathbb{C})$ and the above means there is a natural split
\begin{equation}
    \mathfrak{H}^{2}(\mathbb{R}^{d},\mathbb{C}) =\mathfrak{H}_{-1}(\lambda)\oplus \mathfrak{H}_{\geq 0}(\lambda).
\end{equation}

The set of all such $1$-dimensional subspaces $\mathfrak{H}_{-1}(\lambda)$ of $\mathfrak{H}^{2}(\mathbb{R}^{d},\mathbb{C})$ forms a topological space known as the infinite dimensional Grassmannian $Gr_{1}(\mathfrak{H}^{2}(\mathbb{R}^{d},\mathbb{C}))$ \cite{Grassmannians}, which we denote $Gr_{1}$ for shortness. The space $Gr_{1}$ is a classifying space for the second cohomology group $H^{2}$ \cite{Grassmannians}, \cite{Hatcher-Alg-Top}. This means that (up to homotopy) maps to $Gr_1$ are used to construct the cohomology group $H^2$. Note that the gapped modes condition never entered into our discussion of drift stability. These two conditions are independent as the drift stability is about what happens below $\sigma_{c}(L_{+}(\lambda))$, while the gapped modes condition is about what happens in between (same for gapped Bulk-modes). Thus, we can view linearization around the soliton as a map
\begin{equation}
\Phi^{\lambda} \mapsto \bigg\{\begin{pmatrix}
    0 & P_{\geq0}(\lambda)L_{-}(\lambda)\\
    -P_{\geq 0}(\lambda)L_{+}(\lambda) & 0
    \end{pmatrix}, P_{-1}(\lambda)\bigg\},
\end{equation}
where $P_{-1}(\lambda)$ and $P_{\geq 0}(\lambda$) are projections to $\mathfrak{H}_{-1}(\lambda)$ and $\mathfrak{H}_{\geq 0}(\lambda)$. Because of conditions (\ref{eq:Soliton-potential}), (\ref{eq:Soliton-potential2}), the first component can be seen to be equivalent, up to mode adiabatic evolution (for definition see  sec (\ref{sec:Mode})), to a Hamiltonian operator in $Gap(L^2(\mathbb{R}^{d}))$, the space of gapped $d$-dimensional single-particle periodic Hamiltonians. In \cite{Freed-Moore}, using Bloch's theorem, it is shown using the periodicity that $Gap(L^2(\mathbb{R}^{d}))$ is coarsely equivalent (by adding trivial bands) to $Map(\mathbb{T}^d,BGL_{\infty})$, the space of continuous maps from the $d$-dimensional Brillouin torus to the classifying space $BGL_{\infty}$, which can be thought of as an ever increasing sequence of Grassmannians. Maps (up to homotopy) to $BGL_{\infty}$ give rise to the group $\tilde{\mathit{K}}^{0}$ \cite{Hatcher-K-theory}. Analogously, for systems with a boundary, we instead have $Gap_{Bulk}(L^2(\mathbb{R}^d))$, itself being equivalent to $Map(\mathbb{T}^{d-1},\mathcal{F}^{sa}_{*}(\mathfrak{H}))$ \cite{AASS}, where $\mathbb{T}^{d-1}$ is now the surface Brillouin torus and $\mathcal{F}^{sa}_{*}(\mathfrak{H})$ is a subspace of self-adjoint Fredholm operators \cite{Atiyah-Skew}. Maps to $\mathcal{F}^{sa}_{*}(\mathfrak{H})$ now give rise to the group $\mathit{K}^{-1}$ instead of $\tilde{\mathit{K}}^{0}$. The bulk-boundary correspondence can be seen by noticing that the $-1$ in the degree of the $\mathit{K}$-group compensates the $-1$ in the dimension of the surface Brillouin torus for $d=2$ \cite{Prodan-Schulz-Baldes}. The projection $P_{-1}(\lambda)$ in the second component represents a point in $Gr_{1}$, as discussed previously.

Regarding the scattering of an unstable soliton with a stable soliton, we can speculate that if the unstable soliton implodes, then as the two solitons get closer to each other, it will behave as a strong potential that breaks the gapped modes condition. If the soliton does not implode, however, the expectation is that the other instabilities do not affect these modes, as generally solitons go right through each other under collision. Further non-linear analysis is required. Nevertheless, for solitons sufficiently isolated, these modes should be topologically stable to other perturbations such as defects \cite{Exciton-Polariton}, \cite{Nonlinear-top-photonics}.

\section{Mode adiabatic evolution}\label{sec:Mode}
Consider now eq (\ref{eq:NLS/GP}) with potential, gap energy and non-linearity, which are also $z$-dependent but in such a way that the linearized evolution of the modes around the soliton is adiabatic \cite{Avron-Adiabatic}. We set $1/E_{gap}$ as the adiabatic scale and set $s = E_{gap} z$ to be the dimensionless variable that replaces $z$. We name this key concept \textit{mode adiabatic evolution}. We now use the homotopy interpretation \cite{Avron-Seiler-Simon}, \cite{Freed-Moore} for our mode adiabatic evolution. Two solitons $\Phi^{\lambda_0}(V_0,f_0),\,\Phi^{\lambda_1}(V_1,f_0)$ are in the same topological class if there exists an $s$-dependent family of soliton solutions $\Phi^{\lambda(s)}(V(s),f(s))$ such that

\begin{widetext}
\begin{eqnarray}\label{eq:Mode-Adiabatic}
  [0,1] &\longrightarrow &Gap(L^{2}(\mathbb{R}^{d}))\times Gr_{1} \nonumber\\
  s &\mapsto &
  \bigg\{\begin{pmatrix}
    0 & P_{\geq0}(\lambda(s))L_{-}(\lambda(s))\\
    -P_{\geq 0}(\lambda(s))L_{+}(\lambda(s)) & 0
    \end{pmatrix},P_{-1}(\lambda(s))\bigg\},\nonumber\\
    \Phi^{\lambda(0)}(V(0),f(0)) & = &\Phi^{\lambda_0}(V_0,f_0) ,\nonumber\\
    \Phi^{\lambda(1)}(V(1),f(1)) &=& \Phi^{\lambda_1}(V_1,f_1).
\end{eqnarray}
\end{widetext}

Thus, using this homotopy interpretation of mode adiabatic evolution, we can separate modes into topological classes. Employing the homotopy type of the spaces discussed above, we have that for periodic systems the set of distinct classes of topological modes around solitons is equivalent to the groups:
\begin{equation}\label{eq:local-periodic}
 \tilde{\mathit{K}}^{0}(\mathbb{T}^{d})\oplus H^2(*;\mathbb{Z}),
\end{equation}
 where $*$ denotes (from here on) a point, viewed as a topological space. For systems with a boundary we replace $Gap(L^{2}(\mathbb{R}^{d}))$ with $Gap_{Bulk}(L^2(\mathbb{R}^d))$ and using the results of \cite{AASS}, we obtain $\mathit{K}^{-1}(\mathbb{T}^{d-1})\oplus H^2(*;\mathbb{Z})$ instead. We remark that many solitons of interest such as those that are surface-localized are often gap solitons \cite{Exciton-Polariton} and do not satisfy the spectral condition mentioned above. The topological interpretation of the drift stability condition might seem irrelevant since the group $H^2(*;\mathbb{Z})$ is trivial, but we shall see it yields new classes for systems with more symmetry. Let us reflect on what this result implies. If our solitons satisfy (\ref{eq:Soliton-potential}) and (\ref{eq:Soliton-potential2}), and we ignore issues of soliton stability (momentarily), we can approximate (up to mode adiabatic evolution) eq (\ref{eq:LNLS}) \cite{Weinstein-Zhou} for
\begin{equation}
    \partial_z\chi \approx -i[\mathcal{H}_{l} -\lambda I]\chi\,;\,\, \chi = \chi_1 +i\chi_2,\,\, \begin{pmatrix}\chi_1\\\chi_2\end{pmatrix} = \vec{\chi},
\end{equation}
which is a linear Schr\"odinger equation for the mode $\chi$. This means that all the well-known invariants, such as the Chern number \cite{TKNN}, the spectral flow (for systems with boundary) \cite{AASS}, the $\mathbb{Z}_2$ index (if we add time-reversal symmetry) or mirror Chern number (when we add crystallographic symmetries), etc., have a physical analogue for the mode $\chi$, as discussed for photonic systems in \cite{Top-photonics}, if they exist for the original linear system $\mathcal{H}_{l}$. These invariants can be seen as generating the different $\mathit{K}$-groups which arise. Thus, $\mathit{K}$-groups handle in a single swoop all the different topological mode invariants, instead of discussing each one of them separately.

Let us now switch gears a bit and consider solitons which lie between the gaps of the linear periodic system \cite{Pelinovsky-Oscillatory}, \cite{Top-BandGap-Solitons}, \cite{Weinstein-Periodic}. These are not generally positive and the above spectral condition does not apply. Gap solitons may suffer an oscillatory instability  \cite{Pelinovsky-Oscillatory-Numerical}, \cite{Pelinovsky-Oscillatory}, and the condition for gap solitons to be oscillatory stable can also be stated in topological terms. Let $L(\lambda) = \mathcal{H}_{l} -\lambda I$, the neutral (sometimes called internal) modes (eigenvectors with positive eigenvalues below $\sigma_c (L_{+}(\lambda))$) of $L_{+}(\lambda)$ must lie between the band gaps of the inverted operator $-L(\lambda)$. If they embed in the bands of $-L(\lambda)$, the eigenvalues of $\mathcal{L}(\lambda)$ start bifurcating into complex pairs \cite{Pelinovsky-Oscillatory}. If there are $m$ neutral modes between the inverted bands, then there is an $m$-dimensional vector space $\mathfrak{H}_{N}$ associated to these, which will not change dimension under mode adiabatic evolution, as long as it remains oscillatory stable. Thus, again, we have a natural splitting
\begin{equation}\label{eq:Neutral}
  \mathfrak{H}^{2}(\mathbb{R}^{d},\mathbb{C}) = \mathfrak{H}_{N}(\lambda)\oplus \mathfrak{H}_{rest}(\lambda).  
\end{equation}
 Once again, the space of all possible $m$-dimensional subspaces $\mathfrak{H}_{N}(\lambda)$ of $\mathfrak{H}^{2}(\mathbb{R}^{d},\mathbb{C})$ is the $m$-dimensional Grassmannian $Gr_{m}(\mathfrak{H}^2(\mathbb{R}^{d},\mathbb{C}))$. We can again simplify matters by replacing the finite-dimensional Grassmannian with $BGL_{\infty}$ \cite{Hatcher-K-theory}. Repeating the same analysis as before we obtain the topological classes for oscillatory stable gap solitons, which are given by
\begin{equation}
 \tilde{\mathit{K}}^{0}(\mathbb{T}^{d})\oplus \tilde{\mathit{K}}^{0}(*).
\end{equation}
 We have not seen if this condition generalizes to systems with a boundary; however, considering the appearance of topological surface gap solitons \cite{Exciton-Polariton} and further assuming there is no qualitative difference in their oscillatory behaviour (relative to their bulk analogues), we could suggest that for systems with a boundary, their gap soliton topological classes are given by $\mathit{K}^{-1}(\mathbb{T}^{d-1})\oplus \tilde{\mathit{K}}^{0}(*)$. Once again, the reformulation of oscillatory stability seems irrelevant but let us see what happens when we add symmetries.
 
 On a last note, surface gap solitons can suffer from other types of instabilities such as decaying to small amplitude linear waves. It was found in \cite{Leykam-surface} that solitons with topological edge modes are stable and propagate unidirectionally as in the linear case. It would be interesting to express these results in terms of our mode invariants.
 
 \section{Crystallographic and Time-reversal symmetries}\label{sec:Crystal}
 Consider systems which further have a crystallographic symmetry with point group $P \subset O(d)$ \cite{Fu-Crystalline}, \cite{Freed-Moore}. If we restrict to $P$-symmetric soliton solutions, their corresponding $L_{\pm}(\lambda)$ will be $P$-invariant. Further, if they satisfy all of the conditions discussed above and the mode adiabatic evolution respects this $P$-invariance, then the crystalline topological classes of modes around $P$-symmetric positive solitons are given by:
\begin{equation}
    \bar{\mathit{K}}^{0,\tau}_{P}(\mathbb{T}^{d})\oplus H^{2}_{P}(*;\mathbb{Z}).
\end{equation}
 The groups $\bar{\mathit{K}}^{0}_P$ and $H^2_P$ denote a twisted equivariant version of $\mathit{K}$-theory \footnote{By $\bar{\mathit{K}}^{0}_P(\mathbb{T}^{d})$ we mean the kernel $\mathit{K}^{0}_P(\mathbb{T}^{d})\rightarrow \mathit{K}^{0}(*)$. This is because the coarser classification does not care about adding trivial bands and changing the dimension of our bundles.}, \cite{Freed-Moore} and equivariant cohomology \cite{Gomi-Twists}, \cite{Adem-Groupcoho}, respectively. 
The interesting thing here is that $H^2_{P}(*;\mathbb{Z})$ is no longer trivial! Instead it is equivalent to $H^{2}(BP;\mathbb{Z})$, where $BP$ is an infinite dimensional space known as the classifying space of $P$ \cite{Hatcher-Alg-Top}. To have an example in mind note that for $P = \mathbb{Z}_2$, $B\mathbb{Z}_2 \simeq \mathbb{R}P^{\infty}$, the infinite dimensional real projective space. Hence, the spectral condition becomes topologically non-trivial when we include more symmetries.  For systems with a boundary we replace $\bar{\mathit{K}}^{0,\tau}_{P}(\mathbb{T}^{d})$ with $\mathit{K}^{-1,\tau}_{P}(\mathbb{T}^{d-1})$, where $P$ now denotes surface crystallographic symmetry and $d\geq 2$ \cite{DS-Thesis}. This easily extends to $P$-symmetric oscillatory stable solitons and yields the groups $\bar{\mathit{K}}^{0,\tau}_{P}(\mathbb{T}^{d})\oplus \bar{\mathit{K}}^{0}_{P}(*)$ and $\mathit{K}_{P}^{\tau,-1}(\mathbb{T}^{d-1})\oplus\bar{\mathit{K}}^{0}_{P}(*)$. Once again, $\bar{\mathit{K}}^{0}_{P}(*)$ is not trivial, it is equivalent to the well-known representation ring of $P$, $\bar{R}(P)$ \cite{Tom-Dieck-Compact}.

What is the physical interpretation of these classes? On the one hand for positive solitons, $\mathfrak{H}_{-1}$ is a direction of instability which has to be controlled \cite{Weinstein-Book}. For the perturbed solution to remain at most $\epsilon$-distance from $\Phi^{\lambda}$ at any value of $z$, the initial perturbation needs to be at a distance $\delta(\epsilon,\mathfrak{H}_{-1})$ from $\Phi^{\lambda}$. As we adiabatically evolve $\mathcal{L}(\lambda)(s)$, we would expect an $s$-dependence $\delta(\epsilon,s)$; however, the topological character of $\mathfrak{H}_{-1}$ and $P$-symmetry will mean that $\delta$ is $s$-independent. Thus, how close the initial perturbation has to be to our soliton depends only on the topological action of $P$ on $\mathfrak{H}_{-1}$ and the $\epsilon$ chosen. On the other hand for oscillatory stability, the neutral mode subspace $\mathfrak{H}_{N}(\lambda)$ is a $P$-representation. As discussed in \cite{Soffer-Weinstein-Scattering}, neutral modes are relevant for scattering. In particular, there is a mechanism in which non-linear excited bound states dissipate their energy towards radiation modes and the ground state. The dissipation coefficient $\mathit{\Gamma}$ is a function of the neutral modes ($\mathit{\Gamma} \neq 0$ is the non-linear analogue of Fermi's golden rule). As we mode adiabatically evolve our system, so will $\mathit{\Gamma}(s)$ vary; however, we should always be able to split it $\mathit{\Gamma}(s) = \gamma(\mathfrak{H}_{N}(\lambda))\mathit{\Gamma}_{*}(s)$, with $\gamma(\mathfrak{H}_{N}(\lambda))$ only depending on $\mathfrak{H}_{N}(\lambda)$ as a $P$-representation. Furthermore, we speculate that since these modes must be linearly topologically robust, $\gamma(\mathfrak{H}_{N})$ should dominate and imply a qualitatively slower decay of the modes. We leave the interesting task of explicitly determining $\gamma$ to future work. We remark that both of these new invariants do not arise in linear systems.

We now briefly discuss the inclusion of time-reversal symmetry $\Theta$. Since our systems are bosonic, we only have the so called class AI ($\Theta^2 = I$) in the AZ classification \cite{AZ}. We thus have to replace the $\mathit{K}$-groups with $\mathit{KR}$-groups in the sense of Atiyah \cite{Atiyah-Real}, \cite{Kitaev}. Let us focus on the invariants arising from soliton stability. For drift stability, instead of $H^{2}(*;\mathbb{Z})$, we have the group $H^{2}_{\mathbb{Z}_{2}}(*;\mathbb{Z}(1))$ \cite{D-G-AI}, which is trivial. Similarly for oscillatory stability, instead of $\tilde{\mathit{K}}^{0}(*)$ we have the group $\tilde{\mathit{KR}}^{-6}(*)$ \cite{Kitaev}, which using Bott periodicity (a fundamental result in $\mathit{K}$-theory) turns out to be, again, trivial. Hence, we can conclude that for individual solitons in bosonic systems, time-reversal symmetry $\Theta^2 = I$ does not distinguish between the linear classification \cite{Kitaev}, \cite{AASS}.
\begin{table}[]
\begin{tabular}{cccccc}
\toprule
\multicolumn{1}{p{1cm}}{\centering{$d$}} & \multicolumn{1}{p{1.3cm}}{\centering{Boundary    (y/n)}} & $P$ & $\Theta$     & \multicolumn{1}{p{1.2cm}}{L+DS}        & \multicolumn{1}{p{1.2cm}}{L+OS}    \\
\hline
2   & n       & 0   & 0            & $\mathbb{Z}$                     & $\mathbb{Z}$                   \\
2   & y      & 0   & 0            & $\mathbb{Z}$                     & $\mathbb{Z}$                   \\
2   & n       & pm  & 0            & $\mathbb{Z}^2\oplus\mathbb{Z}_2$ & $\mathbb{Z}^2\oplus\mathbb{Z}$ \\
3   & y      & pm  & 0            & $\mathbb{Z}^3\oplus\mathbb{Z}_2$ & $\mathbb{Z}^3\oplus\mathbb{Z}$ \\
2   & n       & 0   & $\Theta^2=I$ & 0                                & 0                              \\
2   & y      & 0   & $\Theta^2=I$ & 0                                & 0\\ \hline                       
\end{tabular}
\caption{Some examples of topological classes for modes around solitons in $d$-dimensional systems, where $P$ is the crystallographic point group, $\Theta$ is the time-reversal operator, L+DS stands for linear plus drift stable topological classes and L+OS for linear plus oscillatory stable topological classes.}
\label{Table1}
\end{table}
We present a few examples in dimension $d =2,\,3$ with either of these symmetries in Table \ref{Table1}.

\section{Spaces of soliton solutions and Global classes}\label{sec:Moduli}
So far our analysis tells us the different topological character of individual solitons, but does a single soliton define the character of eq (\ref{eq:NLS/GP})? Given $\vec{A}, V$ and $f$ there will generally be many solitons which satisfy conditions (\ref{eq:mode-gap}, \ref{eq:Soliton-potential}, \ref{eq:Soliton-potential2}, \ref{eq:VK}). The set of soliton solutions can be given a topology. Let $M_{D}(E_{gap},\vec{A},V,f),\, M_{O}(E_{gap},\vec{A},V,f)$ be subspaces of soliton solutions, which further satisfy either eq (\ref{eq:VK}, \ref{eq:SpectralCondition}) or eq (\ref{eq:Neutral}) respectively. $M_{D}(E_{gap})$ actually forms a manifold \cite{Weinstein-Zhou}, but to our knowledge not much is known about $M_{O}(E_{gap})$. Let us further identify two solitons in $M_{D}(E_{gap})$ (or $M_{O}(E_{gap})$) as equivalent if they only differ by a translation ($\mathbb{Z}^{d}$), a Galilean boost ($\mathbb{R}^{d}$) or a phase ($\mathit{S}^1$) when these are symmetries of $(\vec{A},V,f)$. Abusing notation we will employ the same symbols for the spaces resulting from the identification. Hence a triple $(\vec{A},V,f)$ induces the maps
\begin{equation}\label{eq:Drift-Soliton-Manifold}
    G^{D}_{\vec{A},V,f}:M_{D}(E_{gap})\longrightarrow Map(\mathbb{T}^{d},BGL_{\infty})\times Gr_{1},
\end{equation}
and
\begin{equation}\label{eq:Oscillatory-Soliton-Manifold}
    G^{O}_{\vec{A},V,f}:M_{O}(E_{gap}) \longrightarrow Map(\mathbb{T}^{d},BGL_{\infty})\times BGL_{\infty}.
\end{equation}
The analogues for systems with boundaries are obtained using $Map(\mathbb{T}^{d-1},\mathcal{F}^{sa}_{*}(\mathfrak{H}))$ instead.
Then, if we mode adiabatically evolve the system $(\vec{A}(s),V(s),f(s))$, these spaces $M_{D}(E_{gap},s),\,M_{O}(E_{gap},s)$ will also change and not necessarily in a continuous fashion. However, as long as they can be deformed into one another via a homotopy, we can use $M_{D}(E_{gap}),\,M_{O}(E_{gap})$ as a characteristic of the entire system and study the maps $G^{D}_{\vec{A},V,f},\,G^{O}_{\vec{A},V,f}$ up to homotopy to define global topological classes. We can also easily extend this to include symmetries. Let us consider the easiest example for $f(s,\Psi) = f(s,|\Psi|^2)$ where $M_{D}(E_{gap})$ is the same as an interval \cite{Weinstein-Zhou} once we have made the proper identifications. This interval is simply the $\lambda$'s which satisfy eqs (\ref{eq:mode-gap}), (\ref{eq:Soliton-potential}) and (\ref{eq:Soliton-potential2}). Then $M_{D}(E_{gap})$ can be contracted to a point and the global drift topological classes for these systems are the same (eq (\ref{eq:local-periodic})) as those for individual solitons. However, suppose that we allow our systems to have $f$'s whose $\Psi$-dependence is not only on its modulus. Then the phase symmetry $\Phi^{\lambda} \mapsto e^{i\theta}\Phi^{\lambda}$ \cite{Weinstein-Zhou} is lost, then we have $M_{D}(E_{gap}) = \mathit{S}^1\times (\lambda_{min},\lambda_{max})$ \cite{Tao-Blog} and the topological classes are equivalent to
\begin{equation}\label{eq:Drift-Circle}
    \bar{\mathit{K}}^{0,\tau}_{P}(\mathit{S}^1\times \mathbb{T}^{d})\oplus H^{2}_{P}(\mathit{S}^{1};\mathbb{Z}),
\end{equation}
or the analogue for systems with boundary. Just as an appetizer we note that for $P =0$, the trivial group, we obtain for fully periodic systems in $d =2$, the group $\tilde{\mathit{K}}^{0}(\mathbb{T}^3) = \mathbb{Z}^3$ and with boundary $\mathit{K}^{-1}(\mathbb{T}^2) = \mathbb{Z}^2$, so the presence of a topologically non-trivial space $M_{D}(E_{gap})$ can add more non-trivial topological classes than there are for individual solitons. We leave equivariant computations for future work. 

\section{Summary and Conclusions}\label{sec:Conclusion}
We have attacked, for the first time (to our knowledge), the problem of assigning topological invariants to non-linear topological systems in photonics, exciton-polaritons and BECs simultaneously \cite{Nonlinear-top-photonics}, \cite{Exciton-Polariton}. Our analysis provides under which conditions the modes around solitons have the same topological character as linear phases do and describes how soliton stability conditions become topologically non-trivial when including crystallographic symmetries, providing new invariants for modes, that have no analogue in linear systems. The same does not happen if we instead have bosonic time-reversal symmetry. Using these constructions together with the space of soliton solutions, we build novel, global invariants for the entire system, providing a partial answer to the problem of classifying non-linear topological systems. General optical systems are non-hermitian as they have gains and losses. Thus, a natural extension of this work is to consider the $\mathit{K}$-theory arising from the point and line gap generalizations for these systems \cite{Shiozaki-Non-hermitian}.  It would also be very interesting to see if other stability conditions for gap solitons have a topological character and to combine them with the non-linear topological nature of vortices, skyrmions and other standard topological solitons \cite{Manton-Solitons}.
\\

We thank O. Antol\'in-Camarena, K. Ramos-Musalem and J. Sheinbaum for useful comments.

\bibliography{apssamp}
\end{document}